\documentclass[prc,twocolumn,letterpaper,10pt,twoside,tightenlines,nofootinbib,showpacs,notitlepage,superscriptaddress]{revtex4-1}
\usepackage{amsmath,amsfonts,amssymb,latexsym}
\usepackage{graphicx}
\usepackage[sort&compress]{natbib}
\usepackage{grffile,epstopdf}
\usepackage{subfigure}
\usepackage{hhline}
\newcommand{\m}{\cdot}

\newcommand{\n}{\nonumber \\}

\newcommand{\llkl}{\left\langle}
\newcommand{\rrkl}{\right\rangle}

\newcommand{\kl}{\left(}
\newcommand{\kr}{\right)}

\newcommand{\dd}{\mathrm{d}}

\newcommand{\del}{\partial}

\usepackage{slashed}
\usepackage{tikz}

\makeindex

\begin{document}

\title{Magnetic field influence on the early time dynamics of heavy-ion collisions}
\author{Moritz Greif}
\email{greif@th.physik.uni-frankfurt.de}
\affiliation{Institut f\"ur Theoretische Physik, Johann Wolfgang Goethe-Universit\"at,
Max-von-Laue-Str.\ 1, D-60438 Frankfurt am Main, Germany}
\author{Carsten Greiner}
\affiliation{Institut f\"ur Theoretische Physik, Johann Wolfgang Goethe-Universit\"at,
Max-von-Laue-Str.\ 1, D-60438 Frankfurt am Main, Germany}
\author{Zhe Xu}
\affiliation{Department of Physics, Tsinghua University, Beijing 100084, China}
\affiliation{Collaborative Innovation Center of Quantum Matter, Beijing, China}
\date{\today }

\begin{abstract}
In high energy heavy-ion collisions the magnetic field is very strong right after the nuclei penetrate each other and a non-equilibrium system of quarks and gluons builds up. Even though quarks might not be very abundant initially, their dynamics must necessarily be influenced by the Lorentz force. Employing the 3+1d partonic cascade BAMPS we show that the circular Larmor movement of the quarks leads to a strong positive anisotropic flow of quarks at very soft transverse momenta. We explore the regions where the effect is visible, and explicitly show how collisions damp the effect. As a possible application we look at photon production from the flowing non-equilibrium medium.  
\end{abstract}

\maketitle



\section{Introduction.}
\label{sec:Intro}
Shortly after the collision of heavy nuclei in experiments at the LHC at CERN or RHIC at BNL the energy density is high enough that a so called quark-gluon plasma (QGP) is formed \cite{Arsene:2004fa,Adcox:2004mh,Back:2004je,Adams:2005dq}. Due to the high velocity and small distances  of the nucleons passing each other at the moment of the collision, the magnetic field for non-central collisions in the center of the reaction at, e.g., top RHIC/LHC energies is very high. Indeed, those fields can be expected to be among the largest field strengths in the universe \cite{Huang:2015oca}.
The fields will decay very fast, even assuming a conducting medium which, by Maxwells equations, slows down the decay of the magnetic flux~\cite{Pu:2016ayh}. Strong enough magnetic fields can generate several novel effects, such as the chiral magnetic effect, the chiral vortical effect~\cite{Kharzeev:2007jp,Kharzeev:2010gr,Kharzeev:2009fn,Fukushima:2008xe}, the chiral separation effect~\cite{Kharzeev:2007jp}, or a chiral magnetic wave \cite{Kharzeev:2010gd,Burnier:2011bf}.
In this article, we turn to a fundamental question which has not gained much attention in literature: will there be directly measurable effects of the electromagnetic Lorentz force in heavy-ion collisions? Early attempts~\cite{Voronyuk:2011jd} to model a hadron gas under the influence of magnetic fields did not show strong effects.

The authors of Ref.~\cite{Gursoy:2014aka} have studied the charge dependent directed flow of pions and protons in a simplified analytic model, taking electric and magnetic fields into account. They find a very small signal, owing mainly to the currents induced by electric fields generated by the fast decaying magnetic fields (Faraday effect). For very low $p_T$ however, they see a strong influence of the magnetic field itself (dubbed as ``Hall'' effect). In particular, the magnetic effect becomes important for the directed flow at transverse momenta $p_T\lesssim 0.25~\mathrm{GeV}$ for RHIC and LHC. In this paper, we will also see the growing influence of the magnetic effect at low $p_T$.


Hydrodynamic calculations including magnetic fields are rare, and still under development~\cite{Roy:2015kma,Inghirami:2016iru}. Recently, it has been found that the directed flow of charm quarks is very sensitive to the magnetic and electric field \cite{Das:2016cwd}.
Furthermore, it has been proposed, that the $J/\Psi$  formation becomes anisotropic which leads, e.g., to a sizable elliptic flow at high transverse momenta~\cite{Guo:2015nsa}. 

We attempt an exploratory study of the early-time non-equilibrium dynamics of deconfined  quarks and gluons including simple parametrizations of an external magnetic field. We find that the quark momenta rotate parallel to the event plane and develop a surprisingly large momentum anisotropy at mid- and forward rapidity for very low transverse momenta. This is roughly reminiscent to the Hall effect. 
The particle velocity stems mainly from the large boosts in beam direction (longitudinal expansion), whereas the magnetic field comes from the charged spectator nucleons. 

We see furthermore that the spectra are also slightly enhanced at early times and we show explicitly how collisions damp both the effect of the flow and the spectra. 

This paper is organized as follows. In Sec.~\ref{sec:model} we explain the model setup, including the magnetic field parametrizations. In Sec.~\ref{sec:toyModel} we show the collisionless result which is purely due to the Larmor movement of quarks in the magnetic field, before we turn in Sec.~\ref{sec:BAMPS} to the result with collisions and conclude in Sec.~\ref{sec:summary}.
\begin{figure}[h!]
	\includegraphics[width=0.95\columnwidth]{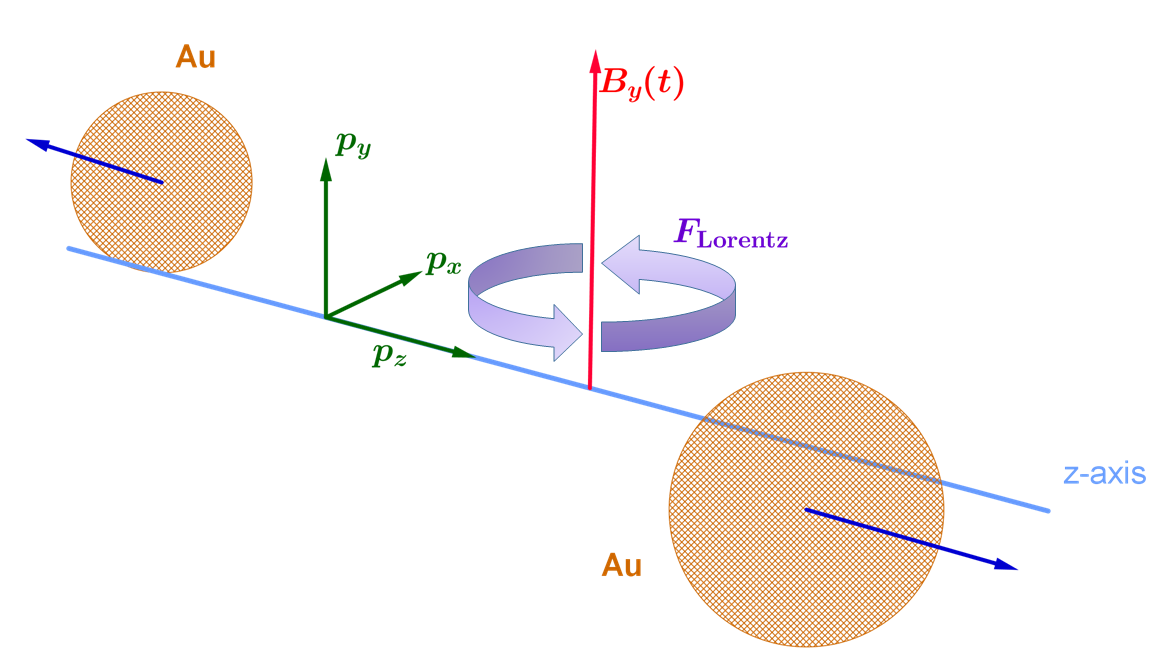}
	\caption{Geometry of our model. The magnetic field is constant and homogeneous.}
	\label{Fig:situation}
\end{figure}

\section{Simple model without collisions}
\label{sec:model}
We investigate how strong the effect of the Lorentz force alone can be on the particle distributions. To this end, we outline a simple model for the heavy-ion collision, neglecting collisions in a first step. We consider the collision of two heavy nuclei along the z-axis. For simplicity, we assume the magnetic field $\vec{B}$ to be constant and homogeneous, pointing in y-direction, $\vec{B}\equiv B_y\vec{e}_y$. The situation is depicted in Fig.~\ref{Fig:situation}. Here we neglect electrodynamical induction effects.
We assume that all events are symmetric and the impact parameter points in x-direction. In this geometry elliptic flow can be seen as an average ($\left\langle \m \right\rangle$) momentum asymmetry $\left\langle (p_x^2-p_y^2)/p_T^2\right\rangle\equiv v_2$, where $p_T=\sqrt{p_x^2+p_y^2}$.

\subsection{Initial state and formation time}
\label{sec:initial_state}
In this simple model setup we do not consider space-dependent effects, thus we sample only four-momenta of the particles. All particles are assumed to be massless. The $p_T$ distribution is sampled according to a power law,
\begin{equation}
\dd N/\dd p_T=\kl \frac{n-1}{p_{T,\mathrm{min}}^{1-n}} \kr p_T^{-n}, \quad n=2,3,4.
\end{equation}
We choose a minimal value $p_{T,\mathrm{min}}=0.01~\mathrm{GeV}$. For all the following results we assume a constant distribution in rapidity, $y=1/2\log(E+p_z)/(E-p_z)$,
\begin{equation}
\dd N/dy=\mathrm{const.},\quad p_z=p_T\sinh y.
\end{equation}
We find that the results are not dependent on the rapidity window in which we initialize the particles, as long as it is larger than the observed rapidity bins. For most studies, $-3<y<3$ is sufficient. 
It is possible to use a formation time $\Delta t_f=\cosh(y)/p_T$ during which particles are still off-shell and do not interact, but propagate freely. This formation time has been used earlier in transport approaches using the Minijet model for the initial condition \cite{Uphoff:2011ad,Eskola:1988yh,Kajantie:1987pd}. 
We can assume that the magnetic field will also not influence the partons within their formation time. However, as quarks carry their electric charge even off-shell, their classical interaction with magnetic fields is arguable, and the formation time could be irrelevant. As this point is conceptionally uncertain, we show results for both options, assuming the particle-field interaction to be switched on immediately (no formation time), or, only after $\Delta t_f$, respectively.
In this simplified collisionless scenario we only initialize quarks, carrying the electric charge $q=e/3$ or $q=2e/3$, respectively. The exact quark and gluon content in the early phase of heavy-ion collisions is under debate. It is clear, that the more gluon dominated the system is, the less pronounced such electromagnetic effects will be.

\subsection{Magnetic field parametrizations}
The external magnetic field present at $t=0^+$ after the collision is still subject to active research, and depends strongly on the geometric modeling of the nuclei as well as the electric conductivity and also possible non-equilibrium effects. 
Common to all the results in literature is the dominant $B_y$-component, perpendicular to the event plane, which is about an order of magnitude larger than the $B_x$-component, the $B_z$-component is nearly absent. The authors of Ref.~\cite{Bloczynski:2012en} look explicitly at fluctuations of the direction of the magnetic field and find that for middle central collisions the field fluctuates less around the direction perpendicular to the event plane than for near central or very peripheral collisions. 
For the qualitative  understanding of the dynamical effects to the quark momenta, we adopt several simplified scenarios for the field strength $B_y$, and set $B_x=B_z=0$. In Ref.~\cite{Voronyuk:2011jd} it was found that the spatial dependence over the overlap region is mild, so that we restrict ourselves here to a homogeneous field in space, parametrized as
\begin{description}
	\item[param 1] $eB_y(t)=4~m_\pi^2\,\Theta(0.3~\mathrm{fm/c}-t)$
	\item[param 2] $eB_y(t)=eB_y^{t=0}(1+t^2/t_c^2)^{-3/2}$ with $t_c=0.065~\mathrm{fm/c}$.
\end{description}
\begin{figure}[h!]
\centering
\includegraphics[width=0.95\columnwidth]{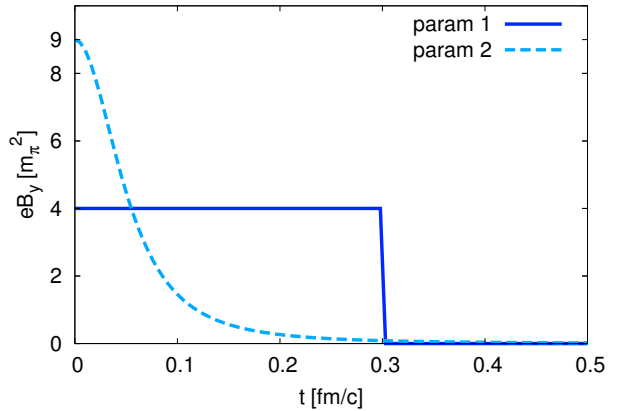}
\caption{The two simple parametrizations of the homogeneous magnetic field. Param 2 follows Ref.~\cite{Deng:2012pc}.}
\label{fig:B_params}
\end{figure}
Param 2 is the parametrization of the results of Ref.~\cite{Deng:2012pc} as given in Ref.~\cite{Huang:2015oca}. We use it with parameters corresponding to RHIC collisions (Au+Au,  $\sqrt{s_{NN}}=200~\mathrm{GeV}$) at typical impact parameters of $\sim 8~\mathrm{fm}$ corresponding to $20-40\%$ centrality (see also, e.g., Refs.~\cite{Bzdak:2011yy,Tuchin:2013apa} for typical field strengths).
In the very early stage, the medium is assumed to be gluon dominated, such that the electric conductivity can be neglected \cite{Huang:2015oca} (being roughly proportional to the sum of the electric charges squared, weighted by the densities of the charge-carrying species \cite{Greif2014,Greif:2016skc}). The authors of Ref.~\cite{Huang:2015oca,Deng:2012pc} approximate the total magnetic field thus by the external component produced by the charged nucleons passing each other.  The full solution of the Maxwell- and Boltzmann equation will slow down the decay of the magnetic field, but so far, only little is known about the precise evolution. Parametrization 1 is an optimistic imitation of a strongly conducting medium, which would keep the magnetic field present for some time. We have tried even higher or longer field parametrizations, but for simplicity we restrict ourselves to an optimistic, and a realistic one.
\subsection{Larmor movement}
The magnetic field changes the direction of velocity of the particles by the Lorentz force, $\vec{F}_{\text{L}}=q\vec{v}\times\vec{B}$. In our geometry, particles will move in a circle around the y-direction, clockwise or anticlockwise depending on their charge $q$. Thereby, any momentum in z-direction will increase or decrease the momentum in x-direction, $p_x\rightarrow p_x+\Delta p_x$.

\begin{figure}[h!]
\centering
\includegraphics[width=0.95\columnwidth]{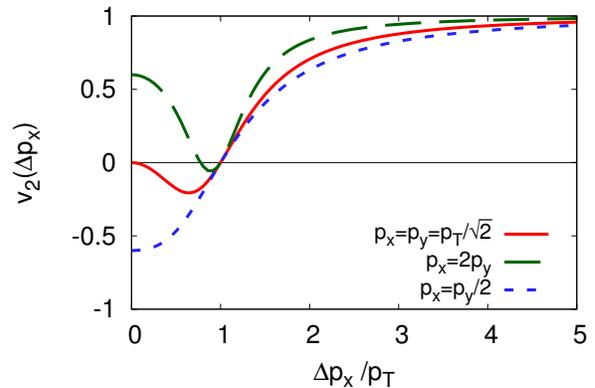}
\caption{After the increase of $p_x$, the momentum asymmetry $v_2$ is in all cases positive for $\Delta p_x>p_T$. The result is symmetric in $\Delta p_x$.}
\label{fig:v2_DeltaPx}
\end{figure}

To analytically estimate the effect of increasing $p_x$ components, we note that by symmetry $\llkl p_x\rrkl = 0,  \llkl p_y \rrkl =0$. We consider two particles with opposite $p_x$ momentum components, $p_{x,1}=-p_{x,2}$ as representer of the particle ensemble. Their $p_y$ momenta are equal, and chosen in a way, that the initial momentum asymmetry $v_2$ takes a given value. The change $p_x\rightarrow p_x+\Delta p_x$ on the $v_2$ of the whole particle ensemble can then, in a simplified fashion, be estimated by
\begin{align}
& v_2(\Delta p_x)\n
&=\frac{1}{2}\kl\frac{(p_x+\Delta p_x)^2-p_y^2}{(p_x+\Delta p_x)^2+p_y^2}+\frac{(-p_x+\Delta p_x)^2-p_y^2}{(-p_x+\Delta p_x)^2+p_y^2}\kr.
\end{align}
In Fig.~\ref{fig:v2_DeltaPx} we show this momentum asymmetry for three choices of the initial $v_2$, positive, zero and negative. Clearly, for zero initial $v_2$, the increase of $\Delta p_x$ must be larger than $p_T$ in order to enhance the asymmetry. All cases show a minimum in $v_2$ for $\Delta p_x<p_T$, which can be strongly negative. 
The radius of deflection due to the magnetic field is
\begin{equation}
r_\text{Larmor}=\frac{	\sqrt{p_x^2+p_z^2}	}{qB_y},
\end{equation}
and the angle of the circular movement of time $t$ is $\alpha_{\text{Larmor}}=t/r_\text{Larmor}$.
The value of $\Delta p_x$ depends on the momenta $p_x$ and $p_z$ of each particle, for a given magnetic field times its duration, $B_y t$.

In this study, we do not include electromagnetic effects other than the Larmor movement. The reason is outlined in the following. The Faraday effect due to the time dependent magnetic flux $\varphi=B_y A$ through surface $A$  generates an electric field, 
\begin{align}
-\frac{\del\varphi}{\del t} = \oint \dd\vec{r}\cdot\vec{E}.
\end{align}
This electric field accelerates charges in the opposite direction than the Lorentz force $\vec{F}_\text{L}=qv\times\vec{B}$. Assuming for a moment, that $\vec{F}_\text{L}\equiv 0$, the electric current due to the force $q\vec{E}$ will generate a magnetic field component $B_{\text{ind}}$ counter balancing the decay of the field, $\vec{B}_{\text{ind}}\sim \del\vec{B}/\del t$, depending on the electric conductivity. On top of these effects, the electric fields generated by the spectators, albeit small in magnitude, has also an $x$-component \cite{Li:2016tel}, which is positive $E_x^{\text{spec}}>0$ for $x>0$ and negative $E_x^{\text{spec}}<0$ for $x<0$.
All these 3 effects can cancel or enhance each other, and depend crucially on the assumed electric conductivity and parametrization of the bare spectator induced fields. Furthermore, the calculation of the magnetic flux as well as the electric fields would require a full space-time dependent (propagating) solution of the electromagnetic fields. This is why we restrict ourselves to show what maximum effect on the particle dynamics is expected from the magnetic field only.  

\section{Results of the collisionless model}
\label{sec:toyModel}
\begin{figure}[t!]
\centering
\includegraphics[width=0.95\columnwidth]{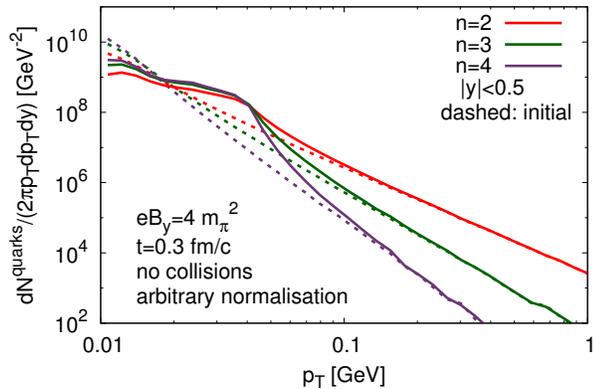}
\caption{For three different initial $p_T$-distributions (power-law exponent $n$) we show how the spectra change after $0.3~\mathrm{fm/c}$ under the influence of a magnetic field. Here we use an arbitrary number of particles and magnetic field parametrization 1.}
\label{fig:spectra}
\end{figure}

\begin{figure}[h!]
\centering
\includegraphics[width=0.95\columnwidth]{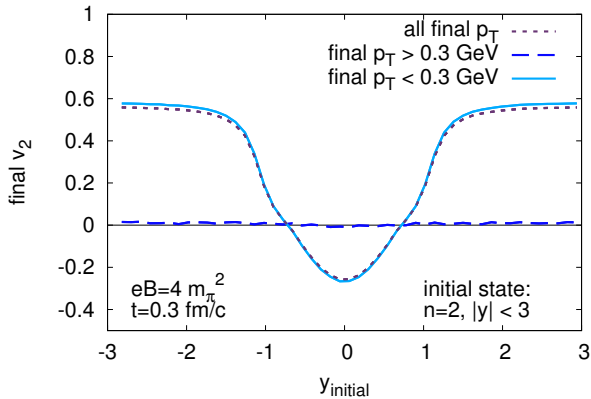}
\caption{Final average $v_2$ per particle for three different $p_T$ ranges as function of initial rapidity $y_{\text{initial}}$ of the particle. Here we use an initial state with power law exponent $n=2$ and magnetic field parametrization 1. The result for $n=3,4$ looks very similar, only the maximal value of the final $v_2$ increases by up to $25\%$.}
\label{fig:v2_y_Mode2}
\end{figure}

First we show how $p_T$-spectra of quarks are influenced in the early time by a magnetic field in Fig.~\ref{fig:spectra}. Here, for parametrization 1 of the magnetic field, the spectrum is enhanced for $0.02<p_T/\mathrm{GeV}<0.1$ due to the Larmor turn of $p_z$-momenta.  
In Fig.~\ref{fig:v2_DeltaPx} we explained that the final momentum asymmetry depends strongly on the additional $\Delta p_x$. 
We explore which momentum space region (regions in rapidity) is necessary to gain sufficiently large values of $\Delta p_x$ for the $v_2$ to change visibly.
Here we differentiate between initial quantities, and those after the circular Larmor-movement has been applied to the particle.
 For this purpose we show in Fig.~\ref{fig:v2_y_Mode2} the final $v_2$ (after the Larmor movement had been applied for a time $t$) as function of initial rapidity $y_{\text{initial}}$. We split this up in a soft region, for final $p_T<0.3~\mathrm{GeV}$, where the averaged $v_2$ reaches large values, and the region of final $p_T>0.3~\mathrm{GeV}$, where the $v_2$ is consistent with zero. This ultrasoft $p_T$ range can already be expected from the spectra, Fig.~\ref{fig:spectra}. Note that the saturation in Fig.~\ref{fig:v2_y_Mode2} is due to the cuts in \textit{final} $p_T$, which means, that, in the curve for final $p_{T}<0.3~\mathrm{GeV}$, the larger $\sinh(y)$, the smaller the values of $p_T$ which contribute.

\begin{figure}
	\centering
	\includegraphics[width=0.95\columnwidth]{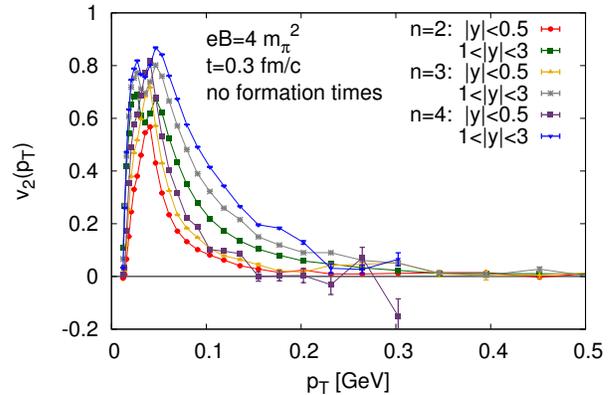}
	\caption{The $p_T$-differential $v_2$ in a collisionless toy model for initial power law spectra with exponent $n=2,3,4$. Here we do not assign formation times. We show results for forward- and midrapidity with magnetic field parametrization 1.}
	\label{fig:v2_pt_woFormationTime}
\end{figure}

\begin{figure}
\centering
\includegraphics[width=0.95\columnwidth]{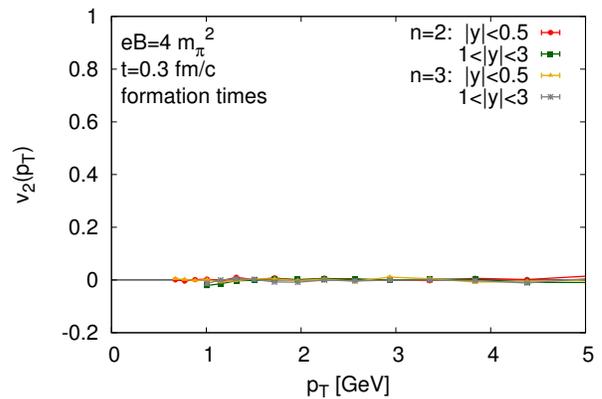}
\caption{Same as Fig.~\ref{fig:v2_pt_woFormationTime}, but here formation times had been assigned. For the shown snapshot at $t=0.3~\mathrm{fm/c}$, the minimum occurring momentum at midrapidity is $p_T\approx 0.66~\mathrm{GeV}$.}
\label{fig:v2_pt_wFormationTime}
\end{figure}

Clearly, momentum rapidities $y>1$ are responsible for $\Delta p_x \gtrsim p_T$ and the average momentum asymmetries larger than zero. The three initial $p_T$-distributions show similar behavior, only the maximal value of $v_2$ increases with increasing $n$.
Finally we turn to the differential $v_2$. Using magnetic field parametrization 1, we show in Fig.~\ref{fig:v2_pt_woFormationTime} the resulting $v_2(p_T)$ without the use of the formation time, for mid- and forward rapidity and all three initial state parametrizations. The $v_2$ can be (temporary) up to $80~\%$. It is larger for forward rapidity. In Fig.~\ref{fig:v2_pt_wFormationTime} we show the result when the formation time was taken into account. This results in deleting all relevant interactions among the field and the particles, and the $v_2$ remains zero.

\section{The effect of collisions}
\begin{figure}[t!]
	\centering
	\includegraphics[width=0.95\columnwidth]{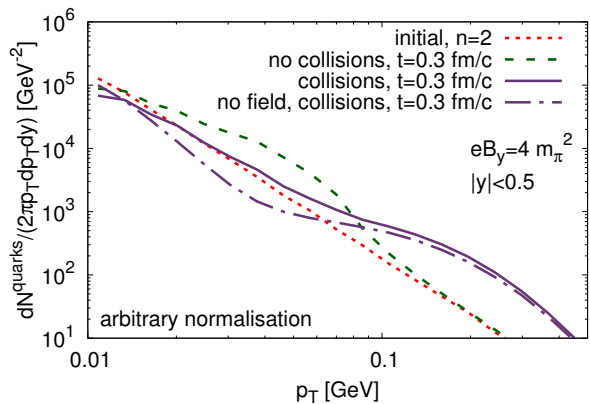}
	\caption{Transverse momentum spectra of quarks under the influence of a magnetic field in a free streaming and a collisional medium. We use an arbitrary number of particles and magnetic field parametrization 1. Particles collide with constant isotropic cross sections, $\sigma_{\text{tot}}=10~\mathrm{mb}$. }
	\label{fig:spectraTimeCollisions}
\end{figure}
\label{sec:BAMPS}
Next we want to consider the effect of particle collisions. To this end we employ the 3+1-dimensional transport approach BAMPS (Boltzmann Approach to Multi-Parton Scatterings), which solves the relativistic Boltzmann equation by Monte-Carlo techniques \cite{Xu2005,Xu:2007aa} for massless on-shell quarks and gluons\footnote{corresponding to an ideal equation of state}. 
The Boltzmann equation is ideally suited to study thermalization and isotropization processes~\cite{Xu2005,Arnold:2007pg} and the electromagnetic fields enter by an external force term.
With the phase-space distribution function $f^i(x,k)\equiv f_\textbf{k}^i$ for particle species $i$, the Boltzmann equation reads
\begin{equation}
k^{\mu }\frac{\partial }{\partial x^{\mu }}f_{\mathbf{k}}^{i}+k_{\nu
}q_{i}F^{\mu \nu }\frac{\partial }{\partial k^{\mu }}f_{\mathbf{k}%
}^{i}=\sum\limits_{j=1}^{N_{\text{species}}}C_{ij}(x^{\mu },k^{\mu }),
\label{eq:BE}
\end{equation}%
where $C_{ij}$ is the collision term, and $q_i$ the electric charge. The field strength tensor $F^{\mu \nu }=E^{\mu }u^{\nu }-E^{\nu }u^{\mu }-B^{\mu \nu}$, with $B^{\mu 0}=B^{0 \nu}=0$, $B^{ij}=-\epsilon^{ijk}B_k$ and $E^\mu=(0,\vec{E})$, introduces the electromagnetic forces to the charged particles~\cite{Cercignani}. For the BAMPS simulations we include 3 flavors of light quarks, antiquarks and gluons. Space is discretized in small cells with volume $\Delta V$ and particles scatter and propagate within timesteps $\Delta t$. In each cell, the probability for binary/inelastic scattering is
\begin{equation}
P_{22/23}=\frac{\sigma_{\text{tot},22/23}(s)}{N_{\text{test}}}v_{\text{rel}}\frac{\Delta t}{\Delta V},
\label{eq:collProb22}
\end{equation}
where $\sigma_{\text{tot}}(s)$ is the (in general Mandelstam $s$ dependent) total cross section and $v_{\text{rel}}$ the relative velocity. The inelastic backreaction works similar. In the simplest case, we employ constant and isotropic cross sections, however, BAMPS features binary and radiative perturbative Quantum Chromo Dynamic (pQCD) cross sections (see, e.g., Ref.~\cite{Fochler:2013epa,Uphoff:2014cba}) and running coupling $\alpha_s(Q^2)$, which is evaluated at the momentum transfer $Q^2$ of the respective scattering process \cite{Uphoff:2012gb}. For the purpose of this study here, there is no qualitative difference when employing pQCD cross sections, so we restrict ourselves to constant and isotropic scattering.
As a new feature, we include the electromagnetic force, which within the Monte-Carlo framework reduces to the additional change of the particle momenta (for every computational timestep) by 
\begin{equation}
\mathrm{d}\vec{k}_i=\Delta t\, F_\text{Lorenz}=\Delta t\, q_i \kl \vec{E}+\vec{v}\times \vec{B}\kr.
\label{eq:FLorentz}
\end{equation}
As mentioned before, we set $\vec{E}=0$. It is clear that the propagation of fields (by retarded Li\'{e}nard-Wiechert potentials) generated by moving quarks would refine the picture, this will be done in a forthcoming publication. Nevertheless,  electric currents appear by default in BAMPS, and the electric conductivity of the matter is built in naturally~\cite{Greif2014}. 
We use the same initial state in momentum space in BAMPS as in the simple model from Sec.~\ref{sec:model}, and use smooth a Glauber Monte Carlo distribution of particle positions. Here we use an impact parameter of $b=8.5~\mathrm{fm}$. The particle numbers are roughly equal to simulations performed in earlier studies using BAMPS for Au+Au collisions at $\sqrt{s_{NN}}=200~\mathrm{GeV}$~\cite{Uphoff:2012gb,Uphoff:2014hza,Uphoff:2014cba}. Flavors for gluons and quarks ($N_f=3$) are sampled randomly with probabilities $P_{g}=16/52, P_q=36/52$. We note that this setup is certainly rough, but it should suffice for our purpose of an optimistic upper estimate of the ``Hall current'' to the anisotropic flow. 
\begin{figure}[t!]
	\centering
	\includegraphics[width=0.95\columnwidth]{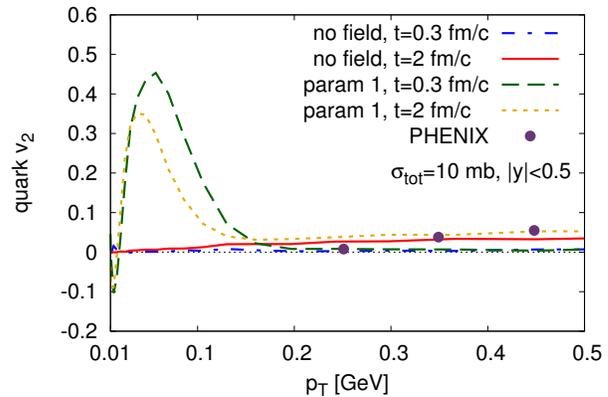}
	\caption{The $p_T$-differential $v_2$ from BAMPS for the ultrasoft $p_T$ range with and without magnetic field. The initial state is equivalent to Sec.~\ref{sec:initial_state}, with exponent $n=2$. The initial geometry is equivalent to an impact parameter of $b=8.5 fm$. We ignore formation times here. For a rough comparison we show data from PHENIX \cite{Afanasiev:2009wq} (unidentified charged hadrons, $\sqrt{s_{NN}}=200~\mathrm{GeV}$, $20\%-60\%$ centrality, $|\eta|<0.35$).}
	\label{fig:v2_pt_lines}
\end{figure}
\begin{figure}[t!]
	\centering
	\includegraphics[width=0.95\columnwidth]{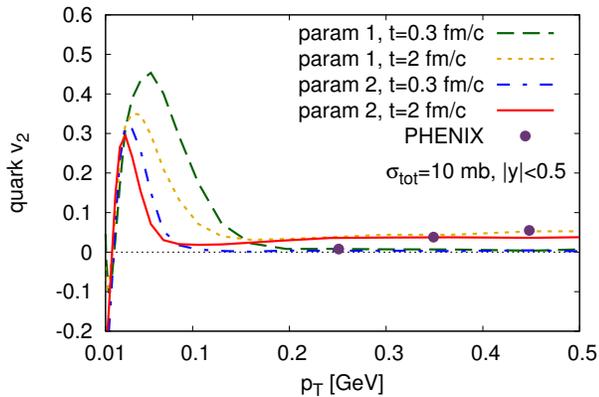}
	\caption{Same as Fig.~\ref{fig:v2_pt_lines}, but here we compare the magnetic field parametrizations 1 and 2.}
	\label{fig:v2_pt_diff_param}
\end{figure}
We show in Fig.~\ref{fig:spectraTimeCollisions}, how the spectra are affected by collisions. Here we see, that in the viscous case (including collisions, $\sigma_{\text{tot}}=10~\mathrm{mb}$), the spectra influenced by the magnetic field for momenta $p_T\gtrsim 1~\mathrm{GeV}$ are very close to the field free case. Without fields, the medium thermalizes at timescales of $0.5\sim 1~\mathrm{fm/c}$ (see Ref.\cite{Xu2005}). Again, in the region of $0.01\lesssim p_T/\mathrm{GeV}\lesssim 0.1$ the spectra are enhanced compared to the field free spectra. The green dashed line shows the collisionless result, which is close to the initial power law for larger momenta, $p_T\gtrsim 1~\mathrm{GeV}$, and enhanced in the soft region.

In Fig.~\ref{fig:v2_pt_lines} we show the $p_T$-differential $v_2$ of light quarks from BAMPS, and turn the magnetic field on and off. Clearly, the field causes a strong momentum anisotropy below $p_T\sim 0.1~\mathrm{GeV}$, but has hardly any effect above this soft $p_T$-range. For comparison we plot the softest points of experimentally measured unidentified charged particle flow from PHENIX \cite{Afanasiev:2009wq}. Unfortunately they are still measured at such high transverse momenta, that a detection of the presented magnetic field effect is unlikely at present.

We see in comparison with Fig.~\ref{fig:v2_pt_woFormationTime}, which shows the collisionless result, that the collisions damp the $v_2$ (about $20\%$ lower $v_2$ at around $p_T=40-60~\mathrm{MeV}$). Here we show results with parametrization 1, which switches off the field at $t=0.3~\mathrm{fm/c}$. After that time, the collisions isotropize this initial flow, such that after $t=2~\mathrm{fm/c}$ it is around $0.34$ and the maximum is pushed to even lower $p_T$.
In Fig.~\ref{fig:v2_pt_diff_param} we compare the effect of the two magnetic field parametrizations. Parametrization 2, probably more realistic, has a weaker effect than parametrization 1. The maximal flow is still about $30\%$, but, more importantly, it is shifted to much lower transverse momenta. We need to recall at this point, that all strong elliptic flow signals appear only, when the formation time of quarks is neglected for the interaction among the field and the particles. This issue must be further addressed in future. 
We note, that we ignore for simplicity hadronization and the subsequent hadron gas evolution here, nevertheless, elliptic flow on the order of $30\%$ is likely to survive to some degree. This remains subject for future work.

Photons are an ideal probe to test effects throughout the spacetime evolution of the medium, such as magnetic field induced flow, as they leave the fireball nearly undisturbed, once produced. In an earlier study \cite{Greif:2016jeb} we have implemented photon production in BAMPS, consistent with leading order rates. Here, we make use of the $2\leftrightarrow 2$ photon production method from Ref.~\cite{Greif:2016jeb}. At very low $p_T$ (where all interesting magnetic effects happen), the microscopic photon production processes for collisions of two low $p_T$ partons will have typical Mandelstam variables at magnitudes, where the concept of perturbative QCD methods is questionable ($s\lesssim \Lambda^2_{\mathrm{QCD}}$). Nevertheless, we allow photons to be produced as we are mainly interested in the non-equilibrium effect of photon production from a flowing quark medium.
 In Fig.~\ref{fig:v2_pt_lines} we show the $v_2(p_T)$ of produced photons with that of the quarks at time $t=2~\mathrm{fm/c}$. Photons are produced during the whole collision, and observables are thus always spacetime averaged, weighted by the production yield. In the beginning of the collision, the medium is dense and the energydensity is high, so many photons are produced, but (in our simplified initial state) the flow is zero. Later, the photons inherit some of the flow, but their rate decreases steadily. This is the reason, why the observed photon flow is smaller than the pure quark flow. Below $p_T\lesssim 0.1~\mathrm{GeV}$ the photon flow is enhanced. It will be challenging to measure such an effect, considering that recent measurements of (direct) photon flow \cite{Adare:2015lcd,Lohner:2012ct_ALICEv2} extend down to $p_T=0.4~\mathrm{GeV}$ (PHENIX)/$p_T=1~\mathrm{GeV}$ (ALICE).

\begin{figure}
\centering
\includegraphics[width=0.95\columnwidth]{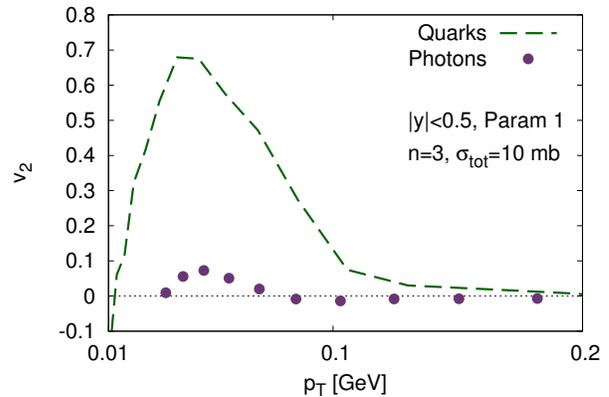}
\caption{Photon $v_2$ as function of $p_T$ compared to quark $v_2$ for magnetic field parametrization 1. Here we use an initial state with power law exponent $n=3$. Photons and quarks are evaluated at midrapidity at $t=2~\mathrm{fm/c}$.}
\label{fig:v2_pt_lines_photons}
\end{figure}

\section{Summary}
\label{sec:summary}
We have shown how the Lorentz force in heavy-ion collisions can affect observables. To this end, we have assumed two simple parametrizations for an external, homogeneous magnetic field, which is produced by the fast spectator nucleons. We investigate a free streaming, and a viscous medium (with collisions), employing the partonic transport simulation BAMPS. We use a simple boost invariant initial state, assuming a power-law in the transverse momentum distribution, and a peripheral Monte-Carlo Glauber geometry of the overlap zone. We have shown that the magnetic field will generate a strong elliptic flow only at very small transverse momenta due to the Larmor movement of the charged particles. In this very soft region, also the transverse momentum spectra are enhanced.  We show that collisions will wash out both the enhancement of the spectra and the elliptic flow. However, the flow is still quite large, such that it could be measured, if experiments had access to ultrasoft transverse momenta.
Assuming an initial formation time of the particles, within which the magnetic field can not act, all strong effects are deleted. The interaction of classical fields and unformed particles is however a difficult theoretical problem and must be clarified further.  
This study can be extended in several ways. Apart from other observables, electric fields, stemming from the Faraday law, might also play a role. A logical next step would be a realistic space dependence of the external fields, and, in the long run, a full spacetime evolution of retarded fields including induction effects (similar to Ref.~\cite{Voronyuk:2011jd}). We emphasize, that the present study should only give an order-of-magnitude estimate of what can be expected from the magnetic Lorentz force (``Hall effect'') for light quarks. Apart of the experimental challenge, there might be other consequences. Especially final spectra of tomographic probes like photons or dileptons will inherit information of this strongly flowing but ultrasoft region. To get an idea of this effect we have shown that photons inherit a fraction of the elliptic flow from the quarks at nearly the same ultralow transverse momenta.

\section*{Acknowledgements}
M.G. is grateful to Tsinghua university in Beijing for their hospitality and acknowledges the support from the ``Helmhotz Graduate School for Heavy Ion research''. The authors are grateful to the Center for Scientific Computing (CSC) Frankfurt for the computing resources. This work was supported by the Helmholtz International Center for FAIR within the framework of the LOEWE program launched by the State of Hesse.
XZ was supported by the MOST, the NSFC under Grants No. 2014CB845400, No. 11275103, No. 11335005, No. 11575092.

\bibliographystyle{apsrev4-1}
\bibliography{library_manuell.bib}

\end{document}